\documentstyle[12pt,aaspp4]{article}
\def\ie{i.e.\ }
\def\eg{e.g.,\ }
\def\etal{et~al.\ }
\def\ltsima{$\; \buildrel < \over \sim \;$}
\def\simlt{\lower.5ex\hbox{\ltsima}}
\def\gtsima{$\; \buildrel > \over \sim \;$}
\def\simgt{\lower.5ex\hbox{\gtsima}}

\def\h1{\ion{H}{1}\ }

\def\h2{H$_2$}
\def\coh2{CO/H$_2$}
\journalid{}{}
\articleid{}{}
\lefthead{Mihos, Spaans, \& McGaugh}
\righthead{The ISM of LSB Disk Galaxies}
\slugcomment{To appear in the {\it Astrophysical Journal}}
 
\begin{document}
 
\title{The Molecular ISM in Low Surface Brightness Disk Galaxies}
 
\author{J. Christopher Mihos,\altaffilmark{1} Marco Spaans,\altaffilmark{2,3}
and Stacy S. McGaugh\altaffilmark{4}}

\altaffiltext{1}{Department of Astronomy, Case Western Reserve University,
10900 Euclid Ave, Cleveland, OH 44106, hos@burro.astr.cwru.edu}
\altaffiltext{2}{Hubble Fellow}
\altaffiltext{3}{Harvard Smithsonian Center for Astrophysics,
60 Garden Street, Cambridge, MA 02138, mspaans@cfa.harvard.edu}
\altaffiltext{4}{Department of Astronomy, University of Maryland,
College Park, MD 20742, ssm@astro.umd.edu}
 
\begin{abstract}

We present models for the interstellar medium in disk galaxies.
In particular, we investigate whether the ISM in low surface
brightness galaxies can support a significant fraction of molecular gas 
given their low metallicity and surface density. It is found that the 
abundance and line brightness of CO in LSB galaxies is small and
typically below current observational limits. Still, depending on
physical details of the ISM, the fraction of gas in the form of molecular 
hydrogen can be significant in the inner few kpc of a low surface brightness 
galaxy.  This molecular gas would be at temperatures of $\sim$ 30 -- 50 K, rather
higher than in high surface brightness galaxies.  These results may help
explain the star forming properties and inferred evolutionary history
of LSB galaxies.

\end{abstract}
 
 
\vfil\eject

\section{Introduction}

Low surface brightness (LSB) disk galaxies represent a class of galactic
systems which have experienced very slow evolution since their formation
epoch. Their low surface brightnesses (\gtsima 1 mag/arcsec$^2$ below
the canonical Freeman (1970) value of $\mu_0^B=21.65 \pm 0.3$ mag/arcsec$^2$)
indicate that, over the age of the Universe, their mean stellar birthrate per 
unit area has been significantly lower than that of typical high surface 
brightness (HSB) disks. Their current rate of star formation is similarly
low --- while some \ion{H}{2} regions do exist in LSBs, the global
star formation rate in LSBs is lower by an order of magnitude than 
comparably sized HSBs (McGaugh 1992; Knezek 1993; McGaugh \& Bothun
1994; R\"onnback \& Bergvall 1994; de Blok, van der Hulst \& Bothun 1995;
de Blok 1997). The lack of
significant star formation is reflected in the low metallicities of
LSBs, which are typically \ltsima 1/3 solar (McGaugh 1994;
R\"onnback \& Bergvall 1995; de Blok \& van der Hulst 1998a).  Not 
coincidentally, LSBs are also very gas-rich systems. McGaugh \& de Blok 
(1997) found that the gas mass fraction of galaxy disks correlates strongly 
with surface brightness. In LSBs, as much as 50\% of the disk mass 
is in the form of gas, compared to $\sim$ 10\% at high surface brightnesses. 
Their low surface brightnesses, low star formation rates, low metallicities,
and large gas fractions all argue that LSBs are systems which are forming 
stars much more slowly than their HSB counterparts.

The suppressed rate of star formation in LSB disks must ultimately be connected
to the differing physical conditions of the interstellar medium (ISM)
between LSB and HSB disk 
galaxies. As star formation is presumed to take place in molecular clouds,
the molecular content of LSBs is of particular interest.  In typical HSB
spirals, the mass of molecular gas is comparable to that in neutral 
\ion{H}{1} (\eg Young \& Knezek 1989). The situation in LSBs may be quite 
different --- while several CO surveys of LSBs have been made (\eg Schombert 
\etal 1990 (S90); Knezek 1993; de Blok \& van der Hulst 1998b (dBvdH)), CO 
emission has not been detected in any LSB disk galaxy. If CO emission traces 
molecular gas content in the same way as in normal HSB galaxies, then the upper 
limits on molecular gas in LSBs are typically $M_{H_2}/M_{HI}$ \ltsima 0.1,
and are more severe in a few cases. These 
upper limits have led to the speculation that the low disk surface
densities in LSBs preclude molecular cloud formation and, in turn, inhibit
star formation (\eg S90; van der Hulst \etal 1993, Bothun \etal 1997).
Alternatively, the lack of CO detection may simply reflect the fact that
the \coh2\ conversion factor is not a universal constant, so that perhaps
large quantities of molecular \h2\ exist despite the lack of detected CO 
emission.

Unfortunately, an observational answer to the question of the molecular
content of LSBs is inexorably tied to the \coh2\ conversion factor
and its dependency on environment. For example, Wilson (1995) and
Israel (1997) recently showed that the \coh2\ conversion factor was a strong 
function of metallicity; this dependency raises the upper limits on the 
derived molecular content of LSBs. Nonetheless, even accounting for 
metallicity effects, previous CO surveys should have detected CO
in LSBs if they had $M_{H_2}/M_{HI}$ ratios similar to HSBs.
Other dependencies should also play a role. For example, the local gas
density and temperature can affect \coh2\
(\eg Maloney \& Black 1988, Scoville \& Sanders 1987).
These are in turn affected by the ionizing radiation field and the
density structure (``clumpiness'') of the ISM. In LSBs all these 
factors may well be significantly different than expected for HSBs,
such that the true molecular-to-atomic gas mass ratio ($M_{H_2}/M_{HI}$) 
is only weakly constrained by direct CO measurements.

To explore the ISM properties of LSB galaxies in a manner independent
of the \coh2\ conversion factor, we take a complementary, theoretical 
route towards understanding the molecular content of LSB galaxies. 
We construct models of an inhomogeneous ISM under varying physical 
conditions, spanning a range of disk galaxy types. The models employ a 
Monte Carlo approach to radiative transfer (see Spaans 1996), and explicitly 
solve for the CO emissivity and $M_{H_2}/M_{HI}$ ratio in galactic disks. 
We investigate models on a grid of metallicity, surface brightness, and 
ISM density structure, tracking the changing physical conditions between
LSB and HSB disk galaxies.  In particular, we
address the questions of how much molecular 
\h2\ is expected in LSB disks, and whether the lack of observed CO in LSBs
in fact indicates a lack of molecular gas.

\section{ISM Modeling}
\subsection{Modeling Technique}

The code developed by Spaans (1996) and its extensions as discussed in
Spaans \& van Dishoeck (1997), Spaans \& Norman (1997), and Spaans \& Carollo
(1998) is used to
derive the physical and chemical structure of the ambient ISM in LSBs.
The interested reader is referred to these papers for a detailed description
of the code's structure. The main features can be summarized as follows.

1) For a given metallicity, geometry, global pressure structure and
distribution of illuminating (ultraviolet) sources, the thermal and chemical
balance of the medium is computed in three dimensions. The continuum
(dust attenuation) and line transfer is modeled through a Monte Carlo
method. The self-shielding of H$_2$ and CO and the shielding of CO by H$_2$
absorption lines is explicitly included. The heating processes include
photo-electric emission by large molecules like Polycyclic Aromatic
Hydrocarbons (PAHs) and dust grains (Bakes \& Tielens 1994), cosmic ray
heating, collisional de-excitation of ultraviolet pumped H$_2$, and H$_2$
dissociation heating.
It is assumed that 10\% of the gas phase carbon is incorporated into PAHs.
This yields roughly equal photo-electric heating
contributions from carbonaceous particles larger
and smaller than $10^{-6}$ cm. Generally photo-electric emission dominates
the heating rate unless the visual extinction exceeds 3 mag.
The cooling processes include fine-structure emission of C$^+$, C and O,
rotational line emission of CO and vibrational (v=1-0) H$_2$ emission. All
level populations are computed in statistical equilibrium and the line emission
is again modeled through a Monte Carlo technique.

2) The solutions to the thermal balance equations allow, for a given
hydrodynamic pressure, multiple solutions. These constitute the possible
multi-phase structure of the ISM as first suggested by Field, Goldsmith, \&
Habing (1969). If multiple solutions exist, then one finds from a stability
analysis that there is a $\sim 10^4$ K diffuse medium and a $\sim 50$ K
dense component. It is the density structure derived from these solutions
which couples strongly with the chemical balance of interstellar gas, and
therefore with the amount of molecular gas which is supported by the stellar
radiation field and the ambient pressure of the galaxy. This thermal
stability approach does not incorporate the effects of hydrodynamic phenomena
such as shocks or gravity. The cold component has a typical density of
$\sim 50-300$ cm$^{-3}$ and is representative of diffuse and translucent
clouds in the Milky Way. To allow the inclusion of shocks and gravity in a
phenomenological way, the dense phase is allowed to exhibit inhomogeneities.
That is, the ambient pressure determines the {\it mean} density of this phase,
while gravity as well as shocks drive perturbations in it.

\subsection{Model Parameters and Their Implementation}

To investigate the molecular content of the ISM the following model parameters
are considered: average gas density, the average UV interstellar radiation 
field (ISRF), metallicity, surface density, and ISM density structure.
These parameters are not all independent. To capture the essential
dependencies of the ISM structure on ambient physical conditions the following
scaling relations are adopted.

The HI volume density $n_{\rm HI}$ correlates with HI surface density 
$\Sigma_{HI}$ according to
$$n_{\rm HI} = \Sigma_{HI}/H,\eqno(1)$$
where $H=300$ pc is the scale height of the galaxy model. Using data from
de Blok \etal (1996), one can derive a rough correlation between 
local surface brightness $\mu^B$ and local HI density: 
$$\log \Sigma_{HI} \approx -0.12*\mu^B + 3.6.\eqno(2)$$
With this relationship, the HI surface density and stellar surface brightness 
do not drop off in lockstep; instead, the HI surface density falls off
more slowly.  While this is generally true, it should be emphasized that
this relation is admittedly crude with a lot of real scatter.  The aim
is more to characterize the general behavior of disks to search for
physically meaningful trends rather than to attempt to model specific
individual galaxies.  In global terms, the gas mass fraction of the 
disk increases as surface brightness decreases
such that very low surface brightness disks ($\mu^B_0$ \gtsima 
23) can have half their baryonic mass in the form of gas (McGaugh \&
de Blok 1997), even assuming a trivial amount of molecular gas mass.

The luminosity profiles of disk galaxies (especially LSBs) are generally
exponential,
$$\mu^B(r) = \mu_0^B + 1.086*(r/h),\eqno(3)$$
with scale length $h$ and central surface brightness in B
mags per square arcsecond $\mu_0^B$.
Combining equations (2) and (3), one finds
$${\rm log}\Sigma_{HI} \approx -0.12*\mu^B_0 -0.13*(r/h) + 3.6.\eqno(4)$$
Again, the relationship implies that, as a function of radius, the
HI surface density drops off more slowly than the stellar surface
brightness, reproducing the extended gaseous disks observed in disk galaxies. 
In this parameterization, the gas surface
density is exponential, but with a scale length 3.3 times larger than
that for the stars. While real gas disks are not as well described by
exponentials as the stellar component, we again stress this is merely
a convenient approximation for modeling purposes.  Deviations from
this approximation will alter only details and not the general trends
of interest, and are probably small compared to the uncertainty
in the modeling process.
Because equation (4) describes the HI surface density, while the model
inputs are in terms of total (HI $+ H_2$) gas surface density,
we use an iterative scheme
to arrive at the final model. First we calculate the model assuming
a total surface density given by equation (4). From this initial model, we 
derive the H$_2$ mass profile, then add this profile to the original
HI profile to produce a total gas mass profile. This total profile is
then used as input to calculate a new, consistent ISM model.

To parameterize the strength of the ISRF in our models, we assume that
the ISRF is dominated by the contribution from the stellar populations
in galaxies. Under this assumption, the ISRF scales with surface brightness:
$$I_{UV} = I_{UV}(MW) * 10^{0.4*(\mu_0^B({\rm MW})-\mu_0^B)}, \eqno(2)$$
where $I_{UV}(MW)$ is the strength of the ISRF in the Milky Way given
by Draine (1978), and $\mu_0^B({\rm MW})$ is the central surface brightness
of the Milky Way disk (assumed to be 21 mag arcsec$^{-2}$).
The wavelengths in the UV relevant to our results are between 912 and 1110
\AA\ where lie all the H$_2$ and CO absorption lines which lead to
photo-dissociation of the molecules.  By scaling the UV ISRF with B-band
surface brightness, we are assuming that the spectral shape is
{\it independent\/} of surface brightness.  That is, we assume
that the stellar populations which give rise to the ISRF
do not drastically change as a function of surface brightness. This
assumption is perhaps suspect.  Since there is generally less star formation
in LSB than in HSB galaxies, one might suspect the UV ISRF to be relatively
weaker in LSBs than implied by the difference in B-band surface brightness.
On the other hand, LSBs do tend to be blue, late type galaxies which have
harder spectral shapes in the optical.  So one might equally well expect
this trend to continue into the UV, resulting in the opposite effect:
the difference in B-band surface brightness might overstate that in the UV.
Without strong constraints on the UV properties of
LSBs we choose simply to hold the shape of
the ISRF fixed with optical surface brightness.
If the UV ISFR is relatively greater [less] than we assume, more [fewer]
molecules will be destroyed and so on balance there will
be less [more] gas mass in molecular form.

With the gas density and UV ISRF defined in terms of the disk surface
brightness, we can similarly define a parameter closely akin
to the ionization parameter:
$$\log{U}=\log(I_{UV}/\Sigma_{HI})=-0.28\mu^B_0 + {\rm constant},$$
which essentially measures the number of ionizing photons per atom.
Because of our assumption that the ISRF scales linearly with
surface brightness, while the gas density drops more slowly, LSB
galaxies should have lower values of $U$ than HSBs. If,
however, LSBs have a harder spectral shape than HSBs (due perhaps
to a younger, hotter mean stellar population), this assumption
may underestimate $U$ in LSBs. While we 
use surface brightness as a fundamental input parameter for the
models, we note that with the pseudo-ionization parameter $U$ defined this
way, models with central surface brightnesses 0, 1, 2, and 3 mag 
arcsec$^{-2}$ below that of the Milky Way correspond to values
of $U/U_{MW}$ = 1.0, 0.5, 0.28, and 0.15, respectively.

Finally, we need to characterize the inhomogeneity of the dense phase, if
it is supported, in the models.
This inhomogeneity can be parameterized by {\it choosing} a certain volume
fraction $F$ of the gas in high density clumps with a fixed density contrast
$C$. The size of the clumps is not varied and assumed equal to 2 pc, typical
for translucent clouds in the Milky Way. By investigating a range of density
contrasts, and therefore clump extinction, this somewhat arbitrary length
does not strongly influence the results. We calculate one model (``H'', see
Table 1) which
is completely homogeneous and lacks any density structure, representing
a limiting extreme. Two more models
are explored which have modest amounts of structure (``I1, I2'', with 
small $C$ and large $F$). Finally, the clumpy ISM models (``C1, C2'', large $C$
and small $F$; see Table 1) are chosen to represent our own Galaxy at high ISM
pressure.

With these parameterizations, we are left with three variables
describing the model galaxies: metallicity, ionization parameter,
and ISM clumpiness. We create a grid of models spanning a range of plausible
values: central surface brightness $\mu_0^B = 21 \to 24$, metallicity
$Z/Z_{\sun} = 1 \to 0.1$, and ISM types H (homogeneous, $P\sim 10^3$ K cm$^{-3}$),
I1 and I2 (intermediate, $P\sim 2\times 10^3$ K cm$^{-3}$), and C1 and C2 (clumpy,
$P\sim 10^4$ K cm$^{-3}$). These models thus capture
the properties of both high surface brightness spirals as well as low
surface brightness disks. For each model we calculate the H$_2$ gas mass
fraction as a function of radius, as well as the CO emissivity
and mass averaged gas temperature. From these models, we can
analyze ISM trends with surface brightness and address the question of
molecular gas content in low surface brightness disks.

\section{Results}

\subsection{Molecular Gas Fractions}

Figure 1 shows $\Sigma_{H_2}/\Sigma_{HI}$ as a function of radius for
several characteristic models. Several trends are immediately obvious:
\begin{itemize}
\item At fixed metallicity and ISM structure, lower surface brightness
models have {\it higher} molecular fractions (Figure 1a). Because the number
of ionizing photons per hydrogen atom decreases with decreasing surface 
brightness, the molecules in the low surface brightness models are less apt 
to be dissociated by the background ISRF.
\item  At fixed surface brightness and ISM density structure, models with 
lower metallicity have lower molecular hydrogen gas content (Figure 1b). 
This result is due to the fact that dust grains act as formation sites 
for molecules; lower metallicities mean fewer dust grains to drive molecule 
formation.
\item At fixed surface brightness and metallicity, clumpier ISM models have
higher molecular gas fractions (Figure 1c). In clumpy models, a larger
mass fraction of the gas is found in denser cores, and are shielded
from the background ISRF. Molecules in diffuse ISM models lack this shielding,
and are more easily dissociated by the UV background.
\end{itemize}

How well do these models describe actual disk galaxies? One point of
constraint is provided by the Milky Way ISM. The high surface brightness,
solar metallicity, and clumpy ISM model shows a mean H$_2$/HI mass ratio
$\sim 1$ averaged across the inner scale length of the disk, similar to
that inferred for Milky Way-like Sb galaxies (Young \& Knezek 1989). This 
result is not surprising, since the ISM models were scaled to the ISRF and 
structure of the Milky Way's ISM, but nonetheless it is reassuring that we 
recover the correct physical description for the given model inputs.

Assigning a model to LSB galaxies is not as straightforward. Certainly
LSB disks are lower in metallicity (Webster \etal 1983; McGaugh 1994;
de Blok \& van der Hulst 1998a) than HSB galaxies such as 
the Milky Way. Their reduced surface brightnesses also probably
results in lower ionization parameters, although stellar population
differences may modify this somewhat.
The density structure of the ISM in LSBs is not well determined,
precisely due to the fact that CO measurements have not yielded any
detections. Because of the lowered mass surface density of LSB disks
(de Blok \& McGaugh 1996, 1997), it is likely that the ISM pressures
are too low to support the amount of multiphase structure found in the
Milky Way. Such was the case in hydrodynamical models of LSB galaxies
by Gerritsen \& de Blok (1998), where a multiphase ISM was virtually 
absent. Models H (homogeneous) and I1 and I2 (intermediate) are therefore
likely candidates to describe the density structure of LSB galaxies.

Figure 2 shows the H$_2$/HI mass ratio averaged over the inner disk
scale length as a function of central surface brightness for the entire
grid of models. For metallicities typical of LSBs ($Z/Z_{\sun}$=0.1--0.3),
the models are lower in molecular content than the Milky Way, as 
expected. Interestingly, though, the models are far from being void
of molecular gas; mass fractions of 0.25 -- 0.5 are typical. Again,
the lowered ionization parameter as a function of surface brightness
results in
{\it higher} molecular fractions (at fixed metallicity and ISM structure) 
for lower surface brightness galaxies. In fact, for very low surface 
brightnesses, the molecular 
content can rival that of HSBs if they have any significant degree
of clumpiness in their ISM. However, at such low surface brightnesses,
the ISM pressures are probably too low to support this level of
structure.

Nonetheless, our models suggest that typical LSB galaxies have
molecular contents which are only factors of 2--3 below that of
normal HSB spirals. The CO mass averaged gas temperatures in the molecular
phase are presented in Figure 3 as a function of radius.
It is immediately obvious that the molecular gas in LSBs is by no means very
cold, in contrast with their multi-phase counterparts. Typical temperatures
are around 30--50 K, similar to Spitzer-type HI clouds in our own Milky Way. 
In Figure 4, we show the cumulative H$_2$ gas mass fraction averaged over the
inner scale length as a function
of temperature for a Milky Way-like model ($\mu^B_0 = 21, Z/Z_{\sun} = 1,$
ISM C2), a typical LSB model ($\mu^B_0 = 23, Z/Z_{\sun} = 0.3,$ ISM I1),
and a very low surface brightness model ($\mu^B_0 = 24, Z/Z_{\sun} = 0.1,$ ISM H).
For the Milky Way model, nearly 50\% of the molecular gas is at or below
30 K, compared to 20\% and only a few percent for the typical and
extreme LSB models. Coupled with the decrease in total molecular
content in the LSB models, our calculations suggest that LSBs should
have very small total amounts of {\it cold} molecular gas.

Such high temperatures argue against efficient star formation in LSBs, but 
self-consistent rates of the order of $\sim$ 0.05 M$_\odot$ yr$^{-1}$ appear 
feasible in these low metallicity environments (Norman \& Spaans 1997;
Gerritsen \& de Blok 1998).
This star formation rate is similar to observed star formation rates in LSBs
(McGaugh \& Bothun 1994; R\"onnback \& Bergvall 1994; de Blok \etal 1995).
In conclusion, the lack of detected CO emission in LSBs
does not preclude the presence of modest amounts of
molecular H$_2$ gas. The CO detectability of an LSB depends on both the
CO abundance and excitation in the galaxy; we turn now to predictions
of the CO intensity of LSBs in order to directly compare to searches
for CO emission from LSBs.

\subsection{CO Intensity and the \coh2\ Conversion Factor}

To calculate the CO intensity of the models,
the root mean square velocity of the interstellar clouds, the vertical
velocity dispersion, is taken equal to 10 km s$^{-1}$, a typical value in the
Milky Way and other galaxies. The turbulent velocity width of individual
clouds is assumed equal to 3 km s$^{-1}$, consistent with the observed
correlation between cloud size and line width for the
Milky Way (Maloney \& Black 1988). We calculate the face-on CO intensities 
for our different ISM models, integrated over the inner scale length. 

Figure 5 shows the variation in I(CO), the CO intensity in K km s$^{-1}$, 
as a function of metallicity, surface brightness, and ISM structure.
As with the H$_2$/HI
mass ratio, several trends are immediately apparent: lower metallicity,
higher surface brightnesses, and a more diffuse ISM all act to lower the
CO intensity in the models. All these trends are as expected. Lower 
metallicities mean fewer carbon and oxygen atoms are available to form 
the CO molecule; higher surface brightnesses result in a stronger
ISRF which destroys the CO molecule; and a diffuse ISM is less effective
at shielding the CO molecules against radiative dissociation.

Also plotted on Figure 5 are the observational upper limits to the CO 
intensity of LSB galaxies determined by S90 and dBvdH. If LSBs have 
solar metallicity, these
observations should have detected CO emission. But the subsolar
metallicities of LSBs (McGaugh 1994) result in lowered CO intensities,
making detection difficult. At $Z/Z_{\sun} \sim 0.3$, the CO emission
is only a factor of $\sim 2 - 5$ below the observational limits,
suggesting that deeper CO mapping may in fact reveal the molecular
ISM of moderately metal poor LSBs. However, reducing the metallicity
by another factor of three reduces the CO emission to levels 30 times
fainter than the current observational limits; detecting these LSBs
in CO will be very hard indeed.  This drop in CO emission occurs in
spite of the presence of a fair amount of H$_2$ in the models.

Perhaps most germane to the observational status of molecular gas
in LSB disk galaxies is the conversion factor $X = n({\rm H}_2)/I({\rm CO})$
(in units of $10^{21}$ cm$^{-2}$ (K km s$^{-1})^{-1})$. Figure 6 shows 
this value calculated for the grid of ISM models. As expected,
$X$ shows significant and systematic variation between 
the different models.  At solar metallicities, $X\sim 0.1-1$, spanning 
the ``standard'' value of X derived from Milky Way observations ($\sim$
0.2 -- 0.5; see, \eg Scoville \& Sanders 1987). Because the CO intensity 
scales non-linearly with density, and in a different manner
from the H$_2$ mass, X has a strong dependence on the 
density structure of the ISM. Our models
calculate the properties of the ISM over the inner disk scale length,
averaging over both cloud and inter-cloud regions. As the
ISM becomes more clumpy, X decreases as the CO intensity rises faster
than the H$_2$ mass fraction. The value of X determined in the Milky
Way may therefore be quite different from that applicable to 
galaxies with a more homogeneous ISM.

Aside from the dependence on ISM density structure, there is also
a clear correlation between X and metallicity: as metallicity drops,
the value of X increases. Such a trend has also been seen in observational
data (\eg Wilson 1995; Israel 1997), and in models of low metallicity
clouds (Maloney \& Black 1988). The strength of this trend is still
quite uncertain. Israel (1997) finds a strong dependence on metallicity 
($\partial \log X/ \partial \log Z = -2.7\pm 0.3$), whereas Wilson (1995) derives a weaker 
relationship, $\partial \log X/ \partial \log Z = -0.67\pm 0.1$.
In our models, the relationship is dependent on the ISM phase structure, but 
falls in the range $\partial \log X/ \partial \log Z = -1 {\rm\ to} -2$.
Again, however, it is difficult to directly compare our theoretical 
values with those determined observationally due to the different
physical scales involved. 

Given the strong dependence on metallicity and ISM density structure,
it is clear that use of the standard Milky Way value of X is suspect in 
LSB galaxies. We can instead turn the problem around and ask, given 
our theoretical calculation of X, what are the inferred constraints on the 
molecular gas fraction of LSBs from the CO studies of S90 and dBvdH. If our 
models are correct, X in LSBs may be greater than the ``standard value''
by as much as a factor of 10, significantly raising the upper limits on LSB
molecular gas content. A similar conclusion was reached by dBvdH, who 
explored the consequences of a non-standard value of X. In that study,
a value of X of four times the galactic value was favored, resulting in
upper limits for LSB molecular contents of $M_{H_2}/M_{HI} < 0.25$. 
Our models favor the use of a high value of $X$ for LSBs, and indicate
that the correct value may be even a factor of two higher than that 
favored by dBvdH. If so, the current non-detections of CO in LSBs still 
allow for significant molecular component of the ISM.  More stringent limits 
on the molecular content of LSBs must await deeper CO observations.

\section{Discussion}

Our models indicate that even very low surface brightness galaxies may not be 
completely void of molecular gas -- instead, the ISM may contain 10--20\% 
of molecular gas (and perhaps more, depending on the detailed physical structure
of the ISM). The physical conditions in this gas may be very 
different from the conditions in the molecular ISM of the Milky Way. If
the ISM pressure is extremely low, as might be expected due to the low
surface mass density of LSB disks, the molecular phase of the ISM will
be diffuse and generally warmer than found in Galactic giant molecular
clouds. The warm
temperature is due largely to the lack of shielding from the ISRF
in a diffuse ISM; even a modest multi-phase ISM can self-shield the
molecular gas and lower the gas temperature. However the low
surface densities and star formation rates of LSB galaxies make it
hard to generate and/or sustain such a multiphase ISM (\eg Gerritsen
\& de Blok 1998).

Aside from the explicit dependencies of the models on ISM structure,
ionization parameter, and metallicity, other more implicit model 
dependencies should also be reiterated. Our models assume that
the UV ISRF scales with optical surface brightness. Stellar population
differences between LSB and HSB galaxies are not well-determined,
but the blue colors of LSBs argue that their stellar populations
may be hotter than those of HSBs. If so, we may underestimate the
ISRF in LSBs, thereby overestimating their molecular content. Similarly,
we have modeled an ISM where the neutral gas density increases
continually into the center of the model, whereas many LSBs show
central depressions of gas density. Again, this effect may push us
towards artificially high molecular contents (by underestimating
the ionization parameter). However, the HI mass profiles of LSBs
are varied, so rather than acting as a systematic effect in our
models, the dependency on gas profile is perhaps better viewed as 
a caution against over-interpreting our results as they apply to 
{\it individual} LSBs. A third model dependency worth noting is the
assumption that the dust-to-gas ratio of the galaxies scales linearly 
with metallicity.  One expects something very close to this from
simple considerations of chemical evolution (Edmunds \& Eales 1998), and
such a relationship is supported by
observational data (\eg Issa, MacLaren, \& Wolfendale 1990).
These dependencies are all
tied to the systematic properties of LSBs which remain ill constrained.
Rather than attempting any further iteration on
the models, we leave these effects as a caveat to the ensuing
discussion.

These uncertainties not withstanding, our models may also shed light 
on the lowered efficiency of star formation
in LSB disks. Compared to HSBs, LSB galaxies have a lower fraction
of molecular material from which they can produce stars. In addition,
whatever molecular gas exists, it is in a more diffuse, warmer
state than is typical for molecular material in HSBs.  These warm 
temperatures and low densities act to help stabilize any existing molecular 
clouds against gravitational collapse. Indeed, since the Jeans length 
scales as $\sqrt{T/\rho}$, the size scale for the collapse of ISM 
substructure is quite large in LSBs. The larger size of any unstable 
patches makes them very susceptible to differential shear in the rotating 
disks, so that gravitational collapse and subsequent star formation in the 
ISM of LSBs will be quite difficult. Even in the solid body portion of
the rotation curve, where rotational shear is not a factor, the star
formation rates remain low due to the increased collapse time of low
density structure.

This stability has been parameterized (\eg Quirk 1972; Kennicutt 1989)
in a form very similar to the Toomre Q parameter for the growth of 
axisymmetric modes (Toomre 1964). Under such prescriptions, star formation
occurs when the gas surface density exceeds some critical value:
$\Sigma_{gas} > \alpha \kappa \sigma / 3.36 G$, where $\kappa$ is the
epicyclic frequency of the disk, $\sigma$ the velocity dispersion of
the gas, and $\alpha$ is a constant $\sim 1$. Studies of LSB galaxies
have shown that the HI surface density is generally below this critical
threshold for star formation (van der Hulst \etal 1993). In fact, the
innermost regions of LSB disks are often suppressed in HI; adding
diffuse, undetected H$_2$ increases the gas surface density and may make
LSB galaxies somewhat more susceptible to induced star formation (\eg
Mihos \etal 1997; O'Neil, Bothun, \& Schombert 1998).  However, the required
amount is not very reasonable.  There are some LSBs with star formation at
small radii where the HI gas is sub-critical by a factor of 4 or more
(de Blok, private communication), quite a bit more than can be made up by
molecular gas for reasonable model parameters.  Whether this is a failure of
our models or of the Quirk-Kennicutt criterion (or both) is unclear.

Similar to the local stability criteria, parameters exist to describe
the stability of disks to growing global bar modes. One such
parameterization is the Toomre $X_2$ parameter: 
$X_2 ={ {\kappa^2 R}\over{4\pi G \Sigma_d} }$, where $\Sigma_d$ is the
total disk mass surface density. If $X_2 >> 1$, disks are stable
against $m=2$ perturbations (Toomre 1981). Mihos \etal 1997 showed that 
because of their lowered disk surface density and increased dark matter 
content (relative to HSB disks), LSB galaxies are quite stable against such
induced bar modes. The inclusion of additional disk mass in the form
of molecular ISM reduces this disk stability, but sufficient dark
matter exists in LSB galaxies to make them stable against all but
the strongest perturbations.

We note in passing that the quantity of mass in this (as yet undetected) 
molecular 
ISM is not nearly sufficient to account for all the dark matter in LSB
disks. Even under the dubious assumption of a maximum (stellar) disk,
de Blok \& McGaugh (1997) showed that the mass deficit in the inner
regions of LSB galaxies is quite severe -- significant amounts of dark
matter must exist all the way into the centers of LSB disks. Under
reasonable assumptions for the physical conditions in LSBs, our models
suggest that the molecular ISM can increase the disk surface 
density {\it at most} by \ltsima 50\%. To account for all the mass deficit implied
by the rotation curve fitting of de Blok \& McGaugh (1997), the molecular
ISM would need to be very cold and very clumpy, raising questions of
why LSBs remain stable and how disk star formation is quenched. 

The different evolutionary histories of HSB and LSB galaxies can be
traced to differences in their disk surface densities and in the 
conditions of their ISMs. A plausible evolutionary scenario for
HSB galaxies has been outlined by Spaans \& Norman (1997). In this
scenario, once the proto-HSB gas disk forms, star formation begins
at a retarded rate in the primordial molecular hydrogen ISM. This
star formation generates supernovae and enriches the ISM, leading to
a multiphase ISM that is able to cool and form stars efficiently --
an HSB disk galaxy is born. In contrast, when a proto-LSB forms,
it, too, forms a molecular ISM, but with a smaller molecular mass
fraction and at lower surface density. At these low surface densities,
it is difficult to trigger star formation or form/maintain a multiphase
ISM. As a result, the LSB evolves little from its primordial conditions,
maintaining its low surface brightness and metallicity, and high gas
fraction. 

Under ``critical density'' conditions for star formation,
one might expect some bimodal surface brightness distribution for
disk galaxies, as galaxies will naturally follow one of two alternative
paths depending on their surface density.  There is a claim of
a bimodal surface brightness distribution in one cluster (Tully
\& Verheijen 1997), but this does not appear to be a general property
of field galaxies (de Jong 1996).
Instead, it is more likely that there is a continuum of physical
conditions in disk galaxies driven ultimately by surface density.
Low density environments result in lowered star formation activity
(as $t_{\rm dyn} \sim \rho^{-{1\over 2}}$ even the absence of any critical
density models)
and suppress the formation of a multiphase ISM; 
as surface density increases along a galactic sequence, star formation 
and surface brightness increase, accompanied by a rise in the
amount of complex phase structure (and higher molecular fractions) 
in the ISM. It is through this interplay that galaxy evolution
and ISM processes shape the (cosmological) star formation rate.

\acknowledgements

We thank Erwin de Blok and Greg Bothun for valuable discussions.
M.S. and J.C.M. have been supported by NASA through Hubble Fellowship grants
\#~HF-01101.01-97A and \#~HF-01074.01-94A, respectively, awarded by the 
Space Telescope Science Institute,
which is operated by the Association of Universities for Research in
Astronomy, Inc., for NASA under contract NAS 5-26555.

\clearpage
 
\begin{figure}
\plotone{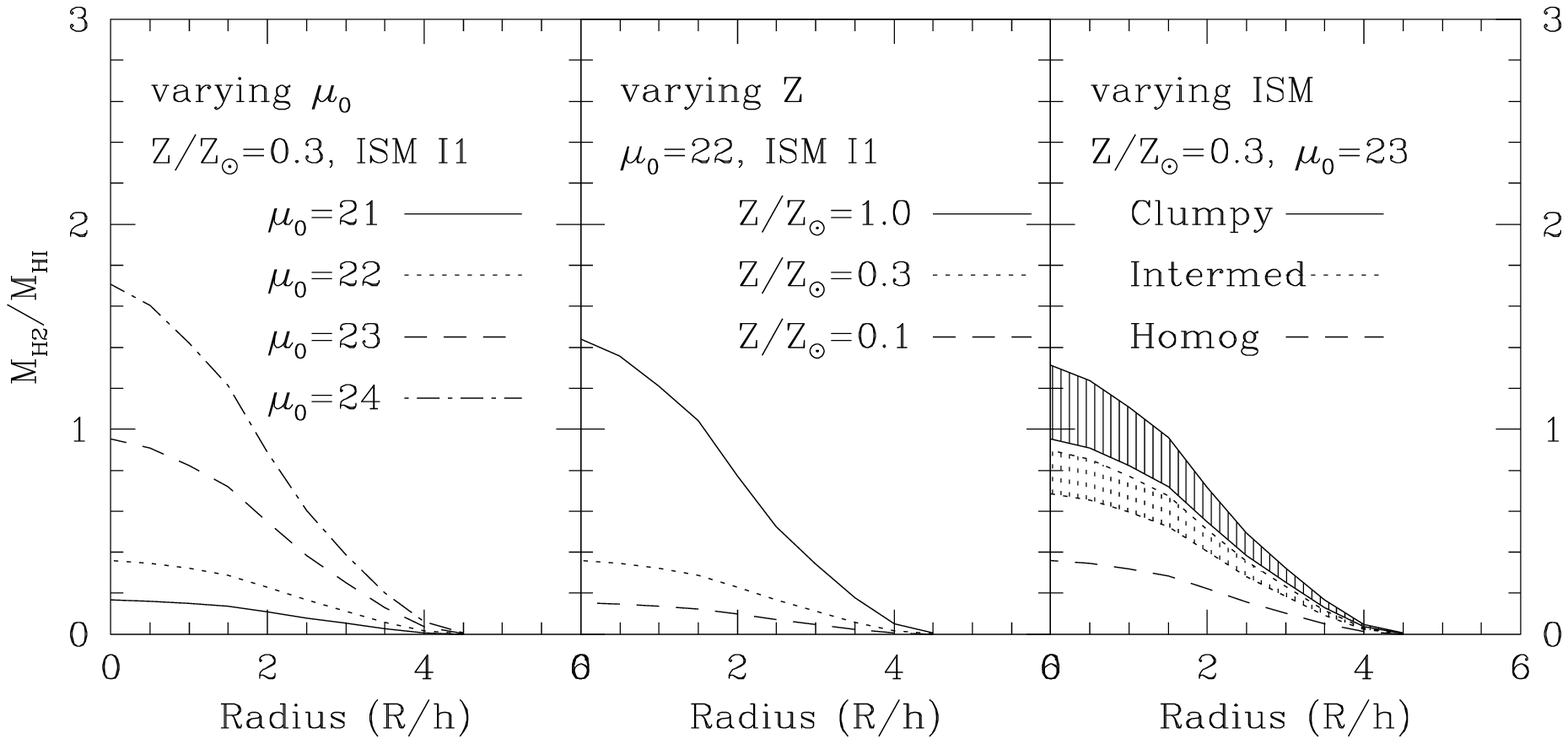}
\caption{The radial H$_2$/HI mass ratio for representative model galaxies.
Radii are measured in units of the disk scale length (\ie $R/h$).
Left: With metallicity and ISM structure held fixed, lowered surface 
brightnesses lead to high molecular mass ratios. Center: With surface
brightness and ISM structure held fixed, the molecular mass ratio drops quickly
with decreasing metallicity. Right: At fixed metallicity and surface brightness,
models with greater degrees of ISM density structure have greater molecular
mass ratios. In this panel, vertical lines connect the two ISM choices for
each of the ``Clumpy'' and ``Intermediate'' ISM models (see text).}
\end{figure}

\begin{figure}
\plotone{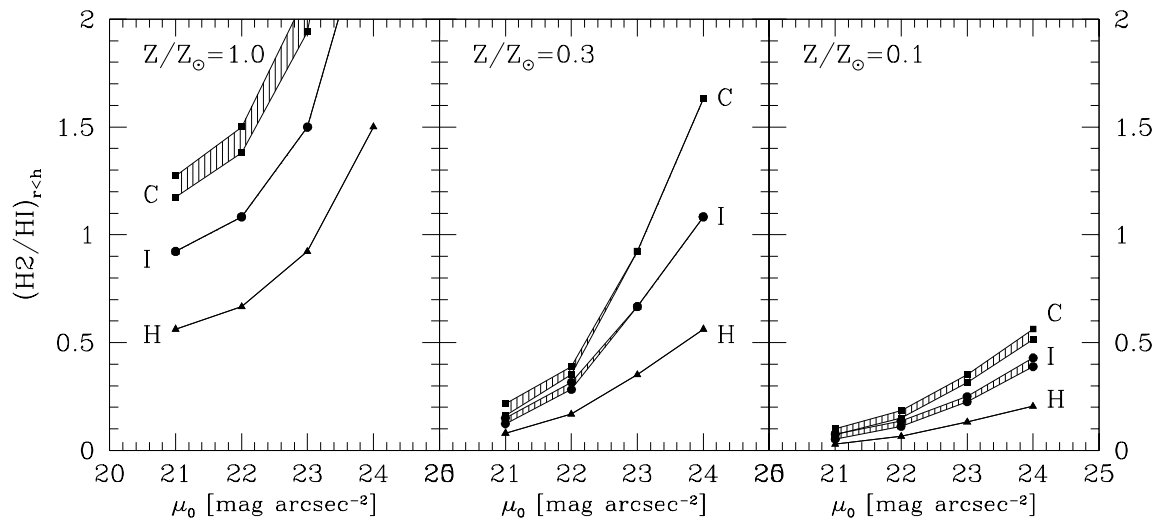}
\caption{The H$_2$/HI mass ratio, averaged over the inner scale length of
the models, as a function of model surface brightness. Curves labeled
``C,'' ``I,'' and ``H'' refer to ISM models with clumpy, intermediate, and 
homogeneous density structure, respectively. Left: Z=Z$_{\sun}$; center:
Z=0.3Z$_{\sun}$; right: Z=0.1Z$_{\sun}$.}
\end{figure}

\begin{figure}
\plotone{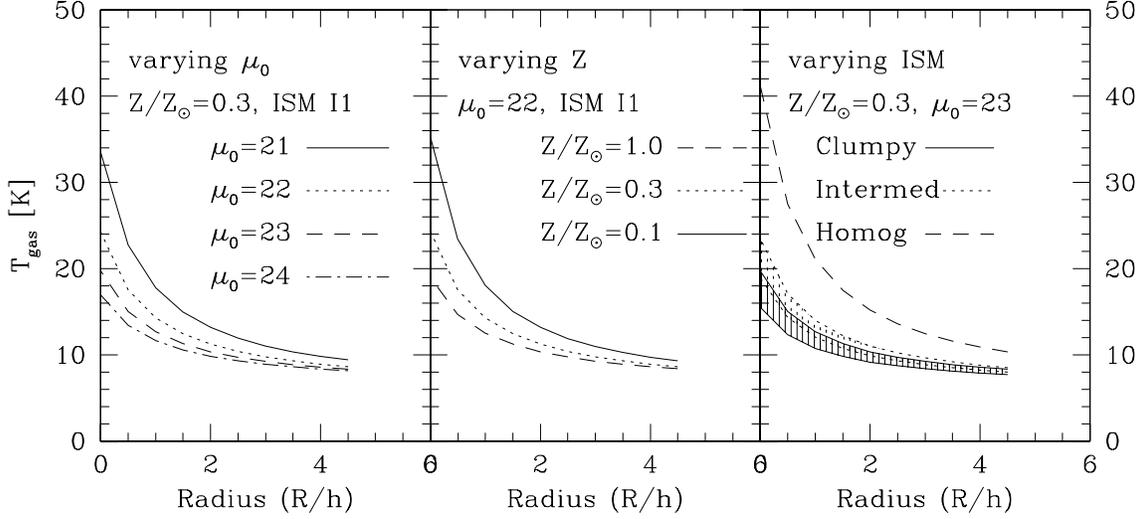}
\caption{Mean molecular gas temperature, as a function of radius, for 
representative galaxy models. 
Radii are measured in units of the disk scale length (\ie $R/h$).
Left: With metallicity and ISM structure held fixed, lowered surface 
brightnesses lead to colder molecular gas. Center: With surface
brightness and ISM structure held fixed, the molecular gas temperature
rises with decreasing metallicity. Right: At fixed metallicity and surface 
brightness, models with less density structure in their ISM have higher
molecular gas temperatures. In this panel, vertical lines connect the two 
ISM choices for
each of the ``Clumpy'' and ``Intermediate'' ISM models (see text).}
\end{figure}

\begin{figure}
\plotone{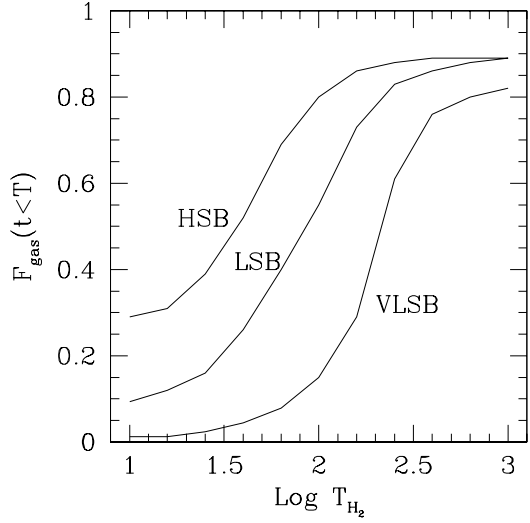}
\caption{A cumulative histogram of molecular gas fraction as a function of
temperature. The high surface brightness (HSB) model has $Z=Z_{\sun}$, 
$\mu_0^B=21$, and the clumpy ``C2'' ISM structure. The low surface
brightness (LSB) model has $Z=0.3Z_{\sun}$, $\mu_0^B=23$, and the intermediate 
``I1'' ISM structure. The very low surface brightness (VLSB) model has 
$Z=0.1Z_{\sun}$, $\mu_0^B=24$, and the homogeneous ``H'' ISM structure.
For the HSB model, nearly half the molecular gas is colder than 30K; for
the LSB and VLSB models only 20\% and \ltsima 5\% of the molecular gas respectively
is that cold.}
\end{figure}

\begin{figure}
\plotone{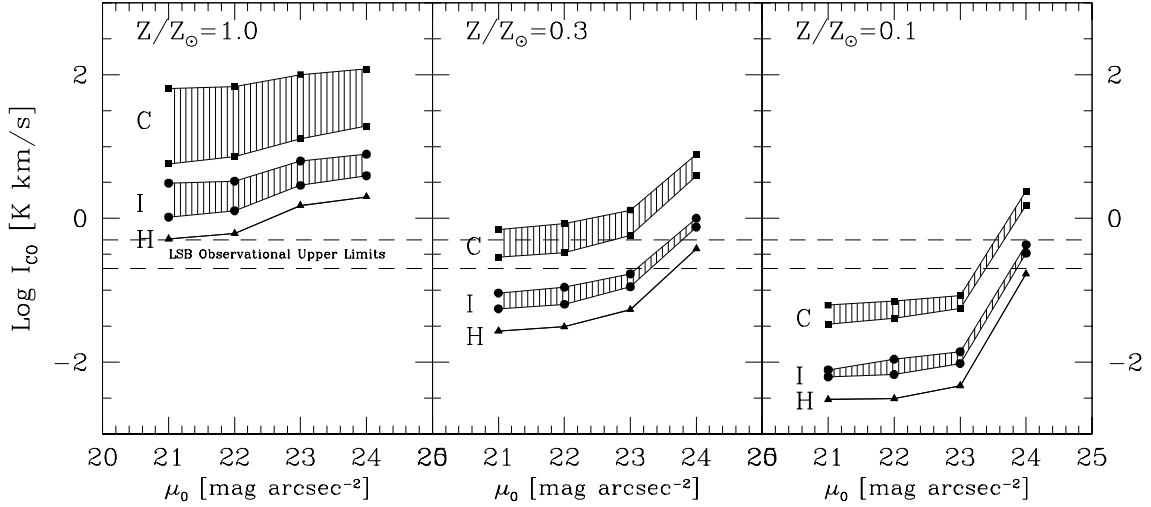}
\caption{The CO intensity as a function of surface brightness for the
different model galaxies. Curves labeled
``C,'' ``I,'' and ``H'' refer to ISM models with clumpy, intermediate, and
homogeneous density structure, respectively. Left: Z=Z$_{\sun}$; center:
Z=0.3Z$_{\sun}$; right: Z=0.1Z$_{\sun}$. The dashed horizontal lines
show the range of observational {\it upper limits} for LSB systems
from S90 and dBvdH. Clumpy, high metallicity systems (such as normal
spiral galaxies) are easy to detect. Low metallicity LSB galaxies have 
much lower CO intensities, but may not be impossible to detect in CO, 
depending on details of their metallicity and ISM density structure.} 
\end{figure}

\begin{figure}
\plotone{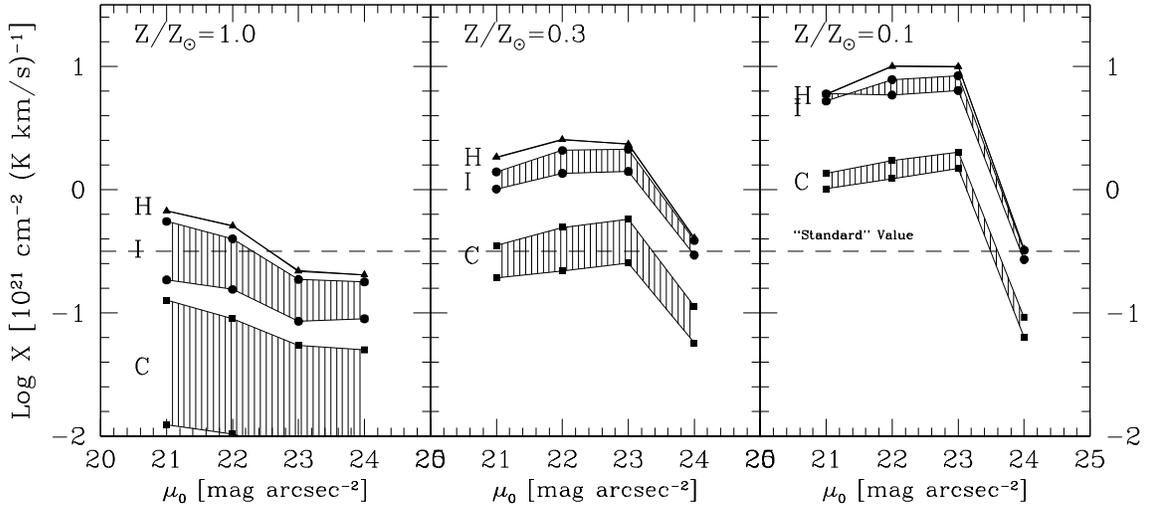}
\caption{The CO-to-H$_2$ conversion factor, $X$, as a function of surface brightness for the
different model galaxies. Curves labeled
``C,'' ``I,'' and ``H'' refer to ISM models with clumpy, intermediate, and
homogeneous density structure, respectively. Left: Z=Z$_{\sun}$; center:
Z=0.3Z$_{\sun}$; right: Z=0.1Z$_{\sun}$. The dashed horizontal line
shows the ``standard'' value typically used in CO studies (\eg Scoville
\& Sanders 1987). Our models show that the value of $X$ has a strong
environmental dependence, varying by as much as two orders of magnitude over the
range of parameters studied.}
\end{figure}

\clearpage

\begin{table}
\caption{ISM Models}
\begin{tabular}{ccc}\hline
Model & C\tablenotemark{a} & F\tablenotemark{b} \\ \hline
H & 1 & 1.0 \\
I1& 2 & 0.5 \\
I2& 4 & 0.25\\
C1& 20& 0.25\\
C2& 60& 0.1 \\ \hline
\end{tabular}
\tablenotetext{a}{Density contrast between high and low density gas.}
\tablenotetext{b}{Volume filling factor of high density gas.}
\end{table}


\clearpage

\begin{references}

\refpar Bakes, E.L.O., \& Tielens, A.G.G.M. 1994, ApJ, 427, 822
\refpar Bothun, G.D., Impey, C., \& McGaugh, S. 1997, \pasp, 109, 745
\refpar Bothun, G.D., Schombert, J.M, Impey, C.D., Sprayberry, D.,
	\& McGaugh, S.S. 1993, \aj, 106, 530
\refpar de Blok, W.J.G. 1997, Ph.D. thesis, University of Groningen
\refpar de Blok, W.J.G., \& McGaugh, S.S. 1996, \apj, 469, L89
\refpar de~Blok, W.J.G., \& McGaugh, S.S. 1997, 290, 533
\refpar de~Blok, W.J.G., McGaugh, S.S., \& van~der~Hulst, J.M.
        1996, \mnras, 283, 18
\refpar de~Blok, W.J.G., \& van~der~Hulst, J.M. 1998a, \aap, 335, 421
\refpar de~Blok, W.J.G., \& van~der~Hulst, J.M. 1998b, \aap, submitted (dBvdH)
\refpar de~Blok, W.J.G., van~der~Hulst, J.~M., \& Bothun, G.D.
        1995, \mnras, 274, 235
\refpar de~Jong, R.S. 1996, \aap, 313, 45
\refpar Draine, B.T. 1978, \apjs, 36, 595
\refpar Edmunds, M.G., \& Eales, S.A. 1998, \mnras, in press
\refpar Field, G.B., Goldsmith, D., \& Habing, H.H. 1969, ApJ, 155, L149
\refpar Freeman, K.C. 1970, ApJ, 160, 811
\refpar Gerritsen, J.P.E., \& de Blok, W.J.G. 1998, A\&A, in press
\refpar Impey, C.D., Sprayberry, D., Irwin, M.J., \& Bothun, G.D. 1996,
	\apjs, 105, 209
\refpar Israel, F.P. 1997, \aap, 328, 471
\refpar Issa, M.R., MacLaren, I, Wolvendale, A.W. 1990, A\&A, 236, 237
\refpar Kennicutt, R.C., Jr. 1989, ApJ, 344, 685
\refpar Knezek, P.M. 1993, Ph.D. thesis, University of Massachusetts
\refpar Larson, R.B. 1982, \mnras, 239, 571
\refpar Maloney, P., \& Black, J.H. 1988, \apj, 325, 389
\refpar McGaugh, S.S. 1992, Ph.D. thesis, University of Michigan
\refpar McGaugh, S.S. 1994, \apj, 426, 135
\refpar McGaugh, S.S. 1996, \mnras, 280, 337
\refpar McGaugh, S.S., \& Bothun, G.D. 1994, \aj, 107, 530
\refpar McGaugh, S.S., \& de Blok, W.J.G. 1997, ApJ, 481, 689
\refpar Mihos, J.C., McGaugh, S.S., \& de Blok, W.J.G. 1997, \apj, 477, L79
\refpar Norman, C.A., \& Spaans, M. 1997, ApJ, 480, 145
\refpar O'Neil, K., Bothun, G.D., \& Schombert, J. 1998, preprint
\refpar Quirk, W.J. 1972, ApJ, 176, L9
\refpar R\"onnback, J., \& Bergvall, N. 1994, A\&AS, 108, 193
\refpar R\"onnback, J., \& Bergvall, N. 1995, \aap, 302, 353
\refpar Scoville, N.Z., \& Sanders, D.B. 1987, in Interstellar Processes,
eds. D.J.Hollenbach \& H.A. Thronson (Dordrecht: Reidel), 21
\refpar Schombert, J.S., Bothun, G.D., Impey, C.D., \& Mundy, L.G. 1990,
	\aj, 100, 1523 (S90)
\refpar Schombert, J.S., Bothun, G.D., Schnieder, S.E., \& 
	McGaugh, S.S. 1992, \aj, 103, 1107
\refpar Spaans, M. 1996, \aap, 307, 271
\refpar Spaans, M., \& Carollo, C.M. 1998, ApJ, 502, 640
\refpar Spaans, M., \& van Dishoeck, E.F. 1997, A\&A, 323, 953
\refpar Spaans, M., \& Norman, C.A. 1997, ApJ, 488, 27
\refpar Sprayberry, D., Bernstein, G.M., Impey, C.~D.,
        \& Bothun, G.~D. 1995, ApJ, 438, 72
\refpar Toomre, A. 1981, in The Structure and Evolution of Normal Galaxies,
Cambridge University Press, p.\ 111
\refpar Toomre, A. 1964, ApJ, 139, 1217
\refpar Tully, R.B., \& Verheijen, M.A.W. 1997, \apj, 484, 145
\refpar van der Hulst, J.M., Skillman, E.D., Smith, T.R.,
        Bothun, G.D., McGaugh, S.S. \& de Blok, W.J.G. 1993, \aj, 106, 548
\refpar Webster, B.L., Longmore, A.J., Hawarden, T.G., \& Mebold, U. 1983,
	\mnras, 205, 643
\refpar Wilson, C.D., 1995, \apj, 448, L97
\refpar Young, J.S., \& Knezek, P.M. 1989, \apj, 347, L55
 
\end{references}
\end{document}